\documentstyle[12pt]{article}
\begin{document}
\font\ninerm = cmr9
% \baselineskip 28pt p
% \pageno=0\footline={\ifnum\pageno>0 \hss --\folio-- \hss \else\fi}

\def\footnoterule{\kern-3pt \hrule width \hsize \kern2.5pt}

\pagestyle{empty}
\begin{center}
{\large\bf Enlarged Bound on the Measurability of Distances
and Quantum $\kappa$-Poincar\`e Group}
\end{center}
\vskip 1.5 cm
\begin{center}
{\bf Giovanni AMELINO-CAMELIA}\\
\end{center}
\begin{center}
{\it Theoretical Physics, University of Oxford,
1 Keble Rd., Oxford OX1 3NP, UK}

\end{center}

\vspace{1cm}
\begin{center}
{\bf ABSTRACT}
\end{center}

{\leftskip=0.6in \rightskip=0.6in

When quantum mechanical 
and general relativistic effects are 
taken into account in the analysis
of distance measurements,
one finds a measurability
bound.
I observe that 
some of the structures
that have been encountered
in the literature on
the Quantum $\kappa$-Poincar\`e Group
naturally lead 
to this bound.

}%%%%%

\vskip 3.4cm

\vfill

\noindent{OUTP-96-54P \space\space\space 
gr-qc/9611016
\hfill August 1996}

\newpage
\baselineskip 12pt plus .5pt minus .5pt
\pagenumbering{arabic}
\pagestyle{plain} 

One of the greatest contemporary challenges for theoretical physics
is posed by 
the incompatibility between
Quantum Mechanics and (classical) General Relativity.
It is likely that the solution of this puzzle, {\it e.g.}
the construction of a quantum theory incorporating gravity,
will require the development of 
a completely new understanding of physics and geometry.
Hints on the structure of the sought new framework
can come from the investigation of problems in which
the incompatibility between
Quantum Mechanics and General Relativity is more evident.
Work in this direction
has led to the expectation that
in Quantum Gravity, unlike ordinary Quantum Mechanics,
there might be bounds on the measurability 
of distances\cite{padma,venezkonish,gacmpla}\footnote{I bring to the
attention of the reader the Refs.\cite{karo,janos,ng}, in which
related issues have been discussed, although
the structures identified in those studies
are {\it gravitational 
corrections}\cite{gacmpla}, rather than measurability bounds.}.

The most commonly\cite{padma,venezkonish} expressed expectation, 
mostly because of its
relevance\cite{venezkonish} 
for the popular critical string theory,
is that there should be a {\it flat} ({\it i.e.} $L$-independent)
bound on the measurability 
of a distance $L$
\begin{eqnarray}
min \left[ \delta L \right] \!\!& = &\!\! L_P
~, \label{qgbounda}
\end{eqnarray}
where $L_P$ is the Planck 
length.
(The distinction between the Planck length and the string length
is inessential to the line of argument here presented.)
Based on the expected inadequacy of ordinary
space-time concepts for scales smaller than 
the Planck length, this can be considered as a minimal 
bound on the measurability of distances in Quantum Gravity.
Within the critical string theory framework it has actually
been possible to find indications\cite{venezkonish}  
that (\ref{qgbounda})
originates from the modified uncertainty relation
\begin{eqnarray}
 \delta x \, \delta P \!\!& = &\!\! \hbar
+ {L_P^2 \over \hbar} \, \delta P^2
~. \label{veneup}
\end{eqnarray}

As discussed in Ref.\cite{gacmpla}, 
an {\it enlarged} (more stringent and $L$-dependent)
measurability bound is suggested by the observation
that, once gravitational effects 
are taken into account, it is no longer possible to rely 
on the availability of {\it classical} agents for the measurement procedure
(the 
limit of infinite masses
leads to inconsistencies\cite{gacmpla} associated with the formation
of horizons).
Based on this observation one arrives\cite{gacmpla}  
at the measurability bound
\begin{eqnarray}
min \left[ \delta L \right] \!\!& = &\!\! 
\sqrt{ {L L_P^2 \over s}}
~,
\label{qgboundgac}
\end{eqnarray}
where $s$ is a length scale characterizing
the spatial 
extension of the devices ({\it e.g.}, clocks) 
used in the measurement\cite{gacmpla}.
This bound is always larger than $L_P$ for acceptable
values\cite{gacmpla} of $s$, {\it i.e.} $L_P \! 
\lower .7ex\hbox{$\;\stackrel{\textstyle <}{\sim}\;$}
 \! s \! 
\lower .7ex\hbox{$\;\stackrel{\textstyle <}{\sim}\;$} \! L$,
and is maximal in the idealized scenario
$s \! \sim \! L_P$, in which
\begin{eqnarray}
min \left[ \delta L \right] \!\!& = &\!\! 
\sqrt{ {L L_P}}
~.
\label{qgboundgacmax}
\end{eqnarray}
A candidate modified uncertainty relation that is based on the general
structure
of (\ref{veneup}), and is motivated by the analysis of
Ref.\cite{gacmpla} (upon disentangling the contributions
of $\delta x$ and $\delta P$ to the final bound),
is given by
\begin{eqnarray}
 \delta x \, \delta P \!\!& = &\!\! \hbar + {c \, L_P^2 \over \hbar} \, 
{T^* \over s} \, \delta P^2
~, \label{gacup}
\end{eqnarray}
where $T^*$ is a time scale characterizing 
the process of observation of 
the system.

In this Letter, I observe that
certain structures
encountered
in the literature on
the Quantum $\kappa$-Poincar\`e Group
naturally lead 
to 
{\it enlarged} measurability bounds of the type (\ref{qgboundgac}),
although the deformed uncertainty relation responsible
for the bound is not of the type (\ref{gacup}). 

I start by  observing that the
quantum $\kappa$-deformed Minkowski space\cite{review,mr,jerzynew}
\begin{eqnarray}
~[x_j , x_k ] \!\!&=&\!\! 0
\label{commrelxx}\\
~[x_j , t ] \!\!&=&\!\! {x_j \over \kappa}
~, \label{commrelxt}
\end{eqnarray}
can be interpreted as implying that
the uncertainties on $x_j$ and $t$
satisfy
\begin{eqnarray}
\delta x_j \, \delta t \!\!&\ge&\!\! {x_j \over |\kappa|}
~. \label{newonegeneral}
\end{eqnarray}
One can find
the implications 
for measurability bounds
of this deformed (obviously, in ordinary Quantum Mechanics 
$\delta x_j \, \delta t \!=\!0$)
uncertainty relation 
by analyzing simple procedures for the measurement 
of the distance $L$
between (the respective centers of mass of) two bodies.
As discussed in Refs.\cite{gacmpla,ng,wign}, 
this type of measurement is naturally carried out by exchanging
a light signal between the two bodies.
Assuming for simplicity
that one of the two bodies 
is a clock,
one can ``attach
non-rigidly''\cite{gacmpla} to it
a ``light gun" ({\it i.e.} a device 
capable of sending
a probe/signal when triggered) and a detector,
and ``attach non-rigidly'' a mirror to the other body. 
The system would be set up so that
a probe be sent toward the mirror
when the clock reads the time $t^{(i)}$, and to record
the time $t^{(f)}$ shown by the clock when
the probe is detected by the detector after being 
reflected by the mirror.
Clearly the time $T \! \equiv \! t^{(f)} \! - \! t^{(i)}$ provides 
a measurement
\footnote{For example,
in Minkowski space and neglecting quantum effects one 
simply finds that
$L \! = \! c T/2$, where $c$ is the speed-of-light constant.
Gravitational (geometrodynamical) effects
would introduce a 
{\it gravitational correction}\cite{gacmpla,karo,janos,ng} $\Delta L$,
but still the measurement of $T$ would result in
a measurement of $L$, based on $L \! = \! \Delta L + T/2$.} 
of the distance $L$.

The relation (\ref{newonegeneral})
could have important implications for the analysis 
of such a measurement procedure.
In fact, it can be interpreted as a relation
between
the uncertainty $\delta t^*$ in the time when the probe 
sets off the detector
and 
the uncertainty $\delta x^*$ in the position
of the probe at the time when it sets off
the detector
\begin{eqnarray}
\delta x^* \, \delta t^* \!\!& \ge &\!\! {2 L \over |\kappa|}
~. \label{newstar}
\end{eqnarray}
Since both $\delta x^*$ and $\delta t^*$ contribute to 
the total uncertainty in the measurement of $L$,
\begin{eqnarray}
[\delta L]_{tot} \ge \delta x^* + c \, \delta t^*
~, \label{newadd}
\end{eqnarray}
the relation (\ref{newstar})
implies that
\begin{eqnarray}
min \left[ \delta L \right] \!\!& \sim &\!\! 
\sqrt{ {c L \over |\kappa|}}
~.
\label{newbound}
\end{eqnarray}
This reproduces the relation (\ref{qgboundgacmax}) 
upon appropriate association
of the scale $\kappa$ 
to the Planck scale.
Actually it is also interesting
to consider $\kappa$ as a second fundamental scale 
(independent of $L_P$)
of Quantum Gravity,
perhaps to be put in relation with $s$ of (\ref{qgboundgac}).

The fact that $\kappa$-Poincar\`e is naturally associated to {\it enlarged}
bounds on the measurability of distances is also reflected in its
generic prediction\cite{review}
of a deformed mass-squared operator, 
which in turn
leads to a 
deformation of the Klein-Gordon equation.
In particular, one of the 
deformations of the Klein-Gordon equation
that have been considered
in the $\kappa$-Poincar\`e literature\cite{review,mr,jerzynew} 
is of the same form of the
deformation of the Klein-Gordon equation
encountered
in the Liouville String investigation reported in Ref.\cite{aemn},
and, 
just as observed in Ref.\cite{aemn},
leads to energy-dependent speeds for massless particles, 
ultimately resulting in a bound
of the type (\ref{qgboundgacmax}) for the measurement
of distances using massless probes\cite{aemn}.

While the case for {\it enlarged}
bounds on the measurability of distances
is certainly made stronger by
the growing list
of candidate Quantum Gravity phenomena
which support them,
much more work needs to be done in order
to get to a satisfactory
mathematical description
of some of the relevant structures.
Perhaps, the most 
prominent of such structures requiring mathematical
work is the ``time of arrival operator''\cite{rovtime}
evoked (more or less explicitly) in 
the analysis here presented.

In the specific context of 
the quantum $\kappa$-deformed Minkowski space
it is also important to 
clarify the physical meaning of the algebraic
concept of noncommutative coordinates\cite{review,connes,wesszumino}.
Once this is clarified,
one might need to reconsider even the statement
that the 
Eqs.(\ref{newonegeneral})
and (\ref{newstar})
follow from
Eq.(\ref{commrelxt}).
Moreover, although
it does not pose a major obstruction for measurement
analysis since 
a preferred
frame
is always identified by
the laboratory ({\it e.g.}, I implicitly assumed above that the center
of the frame coincided with the clock), 
the loss of ordinary translation invariance
encoded in Eq.(\ref{commrelxt})
certainly has deep implications which
would be worth exploring.

The observations here made, when combined with the results 
of Refs.\cite{gacmpla,aemn}, also raise
the possibility of connections between $\kappa$-Poincar\`e
and two other theoretical frameworks.
In fact, the correspondence
between the $\kappa$-analogue of the Klein-Gordon equation
and the deformed Klein-Gordon equation
discussed in Ref.\cite{aemn} 
could be just one aspect of a reacher connection
between the two relevant frameworks, 
{\it e.g.} the geometry underlying Liouville strings\cite{emn}
might turn out to be associated to a
quantum $\kappa$-deformed Minkowski space.
Moreover,
the comparison with the analysis of Ref.\cite{gacmpla}
provides motivation
for the investigation
of the relation between $\kappa$-Poincar\`e and
{\it quantum reference frames}\cite{toller,rovelli,marolf}.

\section*{Acknowledgements}

I am most indebted to J. Lukierski, for very useful
conversations on $\kappa$-Poincar\`e
and feed-back on a preliminary draft
of this Letter.
I also happily aknowledge conversations with
L. Diosi, F. Lizzi, N. Mavromatos, and Y.J. Ng.
This work was supported by funds provided by
the European Union under contract \#ERBCHBGCT940685
within the Human and Capital Mobility program.

\newpage

\baselineskip 12pt plus .5pt minus .5pt

\end{document}